# The Aharonov-Bohm Interference and Beating in Single-Walled Carbon Nanotube Interferometers

Jien Cao, Qian Wang, Marco Rolandi and Hongjie Dai*

*Department of Chemistry and Laboratory for Advanced Materials, Stanford University, Stanford, CA 94305, USA*

Relatively low magnetic fields applied parallel to the axis of a chiral single-walled carbon nanotube are found causing large modulations to the p-channel or valence band conductance of the nanotube in the Fabry-Perot interference regime. Beating in the Aharonov-Bohm type of interference between two field-induced non-degenerate sub-bands of spiraling electrons is responsible for the observed modulation with a pseudo period much smaller than that needed to reach the flux quantum $\Phi_0 = h/e$ through the nanotube cross-section. We show that single-walled nanotubes represent the smallest cylinders exhibiting the Aharonov-Bohm effect with rich interference and beating phenomena arising from well-defined molecular orbitals reflective of the nanotube chirality.

* Email: hdai@stanford.edu

A hallmark of the Aharonov-Bohm (AB) effect [1] is conductance oscillations of metallic rings or cylinders as a function of enclosed magnetic flux with a period on the order of the flux quantum $\Phi_0=h/e$ due to quantum interference [2,3]. Carbon nanotubes are chemically derived cylinders with atomically well-defined structures [4,5]. Multi-walled nanotubes (MWNT) have radius $r \sim 10$ nm and in magnetic fields parallel to the tube axis, conductance modulations with a period of $B_0=\Phi_0 /\pi r^2 \sim 10$T in magnetic field have been seen [6]. Single-walled nanotubes (SWNT) are ultra-small with $r \sim 1$ nm and the magnetic field needed to approach $1\Phi_0$ flux through the nanotube cross section is $B_0 \sim 1000$T, far beyond reach by experiments. We show here that in the Fabry-Perot interference [7] regime, beating in the AB-interference between two modes of spiraling electrons with non-degenerate wave-vectors causes conductance modulations under fields much smaller than that needed to reach $1\Phi_0$.

Ando and Ajiki suggested theoretically that the AB effect manifests in a SWNT by periodically modifying its band structure with a period of $1\Phi_0$ in magnetic flux and $\sim$ 1000 T in field [8]. A periodic change in the band-gap ($\varepsilon_g$) of the nanotube with $d\varepsilon_g/dB \sim$ 1meV/T was predicted. Optical adsorption experiments were able to confirm $d\varepsilon_g/dB \sim$ 1meV/T in semiconducting SWNTs in fields up to 45 T [9]. Recently, electrical measurements also detected similar changes in small band-gap SWNTs transport data inside the band-gap and Coulomb blockade near the band edge [10]. Much can still be done to probe the AB-effect in nanotubes for electronic states far away from the band-gaps and in the ballistic quantum interference regime [7,11] with conductance near $4e^2/h$.

Here, we report the AB effect manifested in the p-channel of a chiral small-gap SWNT in the Fabry-Perot interference regime. We clearly observe beating between two non-degenerate modes of spiraling electrons along the nanotube 'cylinder' while circulating the circumference multiple-turns. Our analysis verifies the predicted band dispersion resulted from the AB effect. Also, by combining single-molecule raman spectroscopy and the AB-interference pattern from electron transport data, we can identify a (15,6) SWNT in the experimental sample.

Our devices comprised of suspended SWNTs grown by chemical vapor deposition (CVD) [12] across trenches (trench height $h$~300nm) between pre-formed Pt contacts [13] (Fig. 1). The suspended tubes are free from substrate perturbations [13] and the high quality of the devices is responsible for the 'clean' experimental data in this work. Micro-raman spectroscopy [14] (Fig. 1c) performed on a $L = 815$ nm long suspended tube (Fig. 1b) revealed three possible assignments to the tube $(m,n) = (14,8)$, (15,6) or (11,11) based on resonance conditions [14,15] with the 785 nm laser used. On the other hand, electrical data suggested a small band-gap nanotube [16,17] due to a gap (with $G$ ~0) in the conductance vs. gate-voltage curve ($G$-$V_g$, Fig. 1d) recorded at T=300 mK. This narrowed down the nanotube to (14,8) or (15,6) with diameter $d$ ~1.5 nm.

The p-channel conductance of the SWNT is high ($\overline{G}$ ~ $2.3e^2/h$) and exhibits an interference pattern in $G$ vs. bias ($V$) and $V_g$, whereas the n-channel shows low conductance and Coulomb blockade (CB) (Fig. 1d). This difference is attributed to lower Schottky barrier to the p-channel of the nanotube than to the n-channel with Pt contacts [13,18]. For the p-channel, the conductance pattern is a result of Fabry-Perot like interference between two degenerate modes (sub-bands) of electrons in the nanotube 'resonator' confined by the two metal contacts [7].

When magnetic fields (-8T to 8T) were applied nearly parallel to the SWNT axis, we observed pronounced conductance modulations (Fig. 2) to the p-channel conductance despite the relatively low fields. The height of the conductance peaks ($\delta G$, relative to valleys) was reduced from $\sim 0.4 e^2/h$ down to $\sim 0.08 e^2/h$ as the field was varied from 0 to 8 T or to $-8$ T (Fig. 2b&c). A slight shift in the positions of the conductance peaks along the $V_g$ axis was also observed as the field increased to 8 T (Fig. 2b&c).

In zero-field, the electron wave-vector $k = (k_\perp, k_\parallel)$ is quantized along the circumference as parallel lines (Fig. 3) such that the wave-function $\Psi(\mathbf{r} + \mathbf{R_{mn}}) = \Psi(\mathbf{r})$ where $R_{mn}$ is the wrapping vector of the nanotube. For a chiral ($m,n$) tube with $m \neq n$ and $m$-$n$=3×integer, two of the $k_\parallel$ lines cross the inequivalent yet degenerate $\mathbf{K_1}$ and $\mathbf{K_2}$ points at the first Brillouin zone corners (Fig. 3a) and give rise to two degenerate sub-bands with zero band-gap. Perturbations such as curvature can cause the zero-gap states deviating from $\mathbf{K_1}$ and $\mathbf{K_2}$ by $\pm \Delta k_\perp{}^0$ respectively (Fig.3a), resulting in a small band-gap for the two sub-bands (two opposite spiralling modes, Fig. 3c) at $\mathbf{K_1}$ and $\mathbf{K_2}$ but maintaining the degeneracy (Fig. 3b) with [8]

$$\varepsilon\left(k_\parallel\right) = \gamma \sqrt{\left(\Delta k_\perp{}^0\right)^2 + k_\parallel^2} \; , \qquad (1)$$

where $\gamma$ is the transfer integral and $2\gamma \Delta k_\perp{}^0 = \varepsilon_g{}^0$ is the band-gap.

In a magnetic field [8], the electron wave-function exhibits a phase shift by $\Psi(\mathbf{r} + \mathbf{R_{mn}}) = \Psi(\mathbf{r})\exp(i\Phi/\Phi_0)$ due to the AB effect. This causes a uniform shift in the allowed states along $k_\perp$ by $\Delta k_{AB} = \dfrac{2\pi}{|\mathbf{R_{mn}}|}\dfrac{\Phi}{\Phi_0}$ (Fig.3d), i.e., shifting the $k_\parallel$ lines for the $\mathbf{K_1}$ and $\mathbf{K_2}$-related sub-bands closer to and further away from the zero-gap states (solid circles in Fig.3a, 3d) respectively. This leads to increased and reduced band-gaps for the two sub-bands respectively and meanwhile lifts their degeneracy (Fig. 3e),

$$\varepsilon\left(k_{\parallel}\right) = \gamma\sqrt{\left(\Delta k_{\perp}^0 \pm \frac{2\pi}{\left|\mathbf{R_{mn}}\right|}\frac{\Phi}{\Phi_0}\right)^2 + k_{\parallel}^2} \quad \text{('+' for } \mathbf{K_1}, \text{ '-' for } \mathbf{K_2} \text{ sub-band)} \qquad (2)$$

The change of bandgap is $d\varepsilon_g / dB \sim \pm\gamma\frac{2\pi}{\left|\mathbf{R_{mn}}\right|}\frac{d\Phi / dB}{\Phi_0} \sim \pm 1$ meV/T. Due to the lifted degeneracy between the $\mathbf{K_1}$ and $\mathbf{K_2}$ related sub-bands by the magnetic field, at a given Fermi energy $\varepsilon$ in the p-channel, two different wave-vector amplitudes now exist (Fig.3e&3f), i.e.,

$$\left|k_{1,2}\right| = \sqrt{\varepsilon^2 - \left(\Delta k_{\perp}^0 \pm \frac{2\pi}{\left|\mathbf{R_{mn}}\right|}\frac{\Phi}{\Phi_0}\right)^2} \quad \text{('+' for } \mathbf{K_1}, \text{ '-' for } \mathbf{K_2} \text{ sub-band)} \qquad (3)$$

for the two modes of electrons with opposite orbiting directions around the nanotube.

We calculated $G$ vs. $B$ and $V_g$ for SWNTs based on interference between non-degenerate $\pm k_1$ and $\pm k_2$ modes (Fig. 4) in a way similar to the Fabry-Perot interference for degenerate modes using the multi-channel Landau-Buttiker formalism and S-matrices [7,19]. Conversion of $\varepsilon$ to $V_g$ was based on matching experimental $G$. vs. $V$ and $V_g$ under $B$=0 (Fig. 1d lower inset) with calculated $G$. vs $V$ and $\varepsilon$ (data not shown). The band-gap of the SWNT $\varepsilon_g^0$ is calculated from ($m,n$) indices based on the curvature induced band-gap model [17]. Numerically calculated G vs. $B$ and $V_g$ for the (15, 6) SWNT give excellent agreement with experimental data (Fig. 2b&2c vs. 4a) in terms of the conductance peaks height modulation vs. $B$ and the amount of peak position shift along $V_g$ under increased field. Calculations based on the (14,8) SWNT do not agree with experiment with much smaller $G$ vs. $B$ modulations than experimental data (Fig.4b vs. Fig.2b&2c).

Up to high fields, simulations reveal that the conductance of the (15,6) SWNT is modulated by $B$ with a pseudo-period of $B_0' \sim 20$-30 T (dependent on $V_g$ or $\varepsilon$) and the conductance peak-shift along $V_g$ becomes more apparent and show 'arching' (Fig. 4a right panel). The experimentally observed $\delta G$ vs. $B$ and $V_g$ well corresponds to such evolutions, albeit in a smaller range of $B$ field. The physics underlying the $G$ modulation with $B_0' \ll B_0$ is beating between two non-degenerate modes of spiraling electrons.

One sees that in a SWNT with greater $|\Delta k_\perp{}^0|$ or larger chiral angle (defined as $\theta$=0 for armchair and 30° for zig-zag tubes), the difference in the number of turns of circumference-orbiting between the two modes when traversing the tube length $L$ is greater than in a tube with zero or small chiral angle. For various $m$-$n$=3×$integer$ SWNTs with similar diameters, beating modulation is the most rapid in zig-zag tubes, followed by chiral tubes and is non-existent in arm-chair SWNTs (Fig. 4c), as confirmed by simulations. By setting the field-induced phase shifts between the two modes over a length of $L$ to $2\pi$, we find an approximate form of $B_0'$,

$$B_0^{'}(\varepsilon) \approx \frac{\pi\ r}{L}\frac{2\varepsilon}{\varepsilon_g^0} B_0 \propto \frac{\pi\ r}{L}\frac{2\varepsilon}{\sin(3\theta)} \cdot B_0 \qquad (4)$$

suggesting that the beating modulation period is reduced from $B_0$ (corresponding to $1\Phi_0$) by a factor of $r/L \sim 10^{-3}$ and is highly sensitive to the tube chiral angle $\theta$ through the $1/\sin(3\theta)$ relation. Note that $\theta$=14° and 9° for (15,6) and (14,8) SWNT respectively, and the difference in chiral angles leads to a large discernable difference in $B_0'$ according to Eq. (4) and simulations (Fig. 4a vs. 4b right panels). For an armchair nanotube, no sub-band splitting occurs due to symmetry and thus no beating-like conductance modulation ($\theta=0$, $B_0' \sim \infty$) by axial magnetic fields (Fig. 4c). Nevertheless, a band-gap is opened for the two degenerate sub-bands and the band-gap change and resulting non-linearity in $\varepsilon(k)$ lead to shifting (or 'arching') (Fig. 4c right panel) of the conductance peaks along $V_g$ under increasing $B$. This is the regular AB effect (in the absence of beating) with a $1\,\Phi_0$ period in magnetic flux and is universal for nanotubes of all chirality (Fig. 4a,b and c right panels).

The observation of quantum beats for the Aharonov-Bohm effect is to our knowledge unprecedented in mesoscopic systems and is a result of well-defined molecular orbitals of nanotubes in magnetic fields. Large band-gap semiconductor SWNTs with low Schottky-barrier p-channels in the Fabry-Perot regime [18] could exhibit much more rapid beats than the small band-gap SWNTs. Clearly, many future opportunities exist for elucidating quantum interference and beating between well-defined molecular orbitals.

We thank Ali Javey for discussions. This work is supported by the MARCO MSD Focus Center, SRC/AMD, INMP, a Packard Fellowship and a Dreyfus Teacher-Scholar.

**Figure Captions:**

**Figure 1**. A suspended chiral-nanotube quantum wire. (a) Schematic device structure. Nanotubes were synthesized across Pt electrodes over trenches at 800-820°C to produce SWNTs with diameters $d$ < 2 nm. (b) Electron micrographs of the device layout (left image) and actual suspended nanotube (right image, $L \sim 815$ nm) used for this work. (c) A resonance micro-raman spectrum (Renishaw, laser $\lambda$=785nm, spot size 1μm scanned over the trench) showing the radial breathing mode (RBM) of a $d \sim$1.5 nm SWNT with possible chirality assignments of (11,11), (14,8) and (15,6) based on the RBM shift at $\varpi$=163 ± 4cm$^{-1}$. (d) $G$-$V_g$ characteristics recorded at T=300mK in a $^3$He cryostat under bias $V$ =1mV. Left inset: $G$ (represented by color, dark blue: ~2$e^2$/$h$, bright white: ~ 2.4 $e^2$/$h$) vs. bias $V$ and $V_g$ for the p-channel showing Fabry-Perot interference pattern. Right inset: $G$ vs. $V$ and $V_g$ for the n-channel displaying Coulomb blockade diamonds.

**Figure 2**. Experimental data of a nanotube interferometer in magnetic fields. (a) $G$-$V_g$ characteristics for the suspended SWNT in magnetic fields (angle to tube axis ~ 9°) from 0 to 8 T recorded at T=300 mK under V =1mV. The conductance peaks monotonically decrease from 0 to 8 T. (b) A zoom-in view of (a). From top to bottom curves, $G$-$V_g$ characteristics in field $B$=0 to 8T, in 2 T steps. Notice slight shifts in the peak positions to the left at higher fields. (c) A plot of $G$ (represented by color) vs. $V_g$ and magnetic field $B$ (-8 T to 8T) based on 160 $G$-$V_g$ curves from $B$ = –8T to 8T in 0.1 T steps. Color scale bar unit: $e^2$/$h$. The slight shifts of conductance peaks positions in (b) are reflected in the slight arching of the interference stripes as highlighted by the dashed line.

**Figure 3**. The AB effect in a multi-mode nanotube interferometer. (a) The first Brillouin zone of a small-gap chiral SWNT (in zero magnetic field). Only two parallel lines (dashed) of the allowed states closest to the zero band-gap points (the two solid circles near the inequivalent $\mathbf{K_1}$ and $\mathbf{K_2}$ corner points respectively) are shown. Perturbations cause the zero band-gap states deviating from $\mathbf{K_1}$ and $\mathbf{K_2}$ by $\pm\Delta k_\perp^0$ due to symmetry and the opening of a small-gap along the two dashed lines ($\mathbf{K_1}$ and $\mathbf{K_2}$ sub-bands). (b) Dispersion $\varepsilon$(k) relations for the two degenerate sub-bands near $\mathbf{K_1}$ (blue curve) and $\mathbf{K_2}$ (red curve) respectively. $k_\parallel$=0 is defined for the states where the conduction and valence band are the closest. (c) Two degenerate modes of spiraling electrons in zero-field. (d) In a magnetic field parallel to the tube axis, the electronic states (dashed lines) shift by $\Delta k_{AB}$ due to the AB effect, leading to lifting of the degeneracy between the two sub-bands. (e) Dispersion curves for the two non-degenerate sub-bands in a magnetic field of $B$=8T. (f) Two non-degenerate modes of spiraling electrons in a magnetic field.

**Figure 4**. Simulation of Aharonov-Bohm interference and beating versus nanotube chirality. (a), (b) and (c) are simulation results for (15,6), (14,8) and (11,11) SWNT respectively. Left panel: Calculated $G$-$V_g$ curves in fields of $B$=0 to 8T from top to bottom in 2T steps. Middle panel: Calculated 2-D plot of $G$ vs. $V_g$ and $B$ in a small field range -8T to 8T. Right panel: Calculated 2-D plot of $G$ vs. $V_g$ and $B$ over a wider field range of -30T to 30 T showing that shifting or arching of the conductance peaks is universal for the three different chirality nanotubes. Our simulations here used precise $\varepsilon(k)$ dispersions such as Eq. 2 and 3 instead of linear approximations. Note that we have also considered and excluded the possibility of a (16,4) tube in the experimental sample. The (16,4) tube has a chiral angle of 19º, larger than the three tubes considered in this figure and exhibit more rapid beating period than the experimental data.  Further, the (16,4) tube has $d \approx 1.43$ nm, outside of the range of $d$=1.45-1.53 nm determined from the Raman resonance condition[15] of $d$ (in nm)=223.5/($\varpi$-12.5) for $\varpi$=163 ± 4cm$^{-1}$ (Fig. 1c).

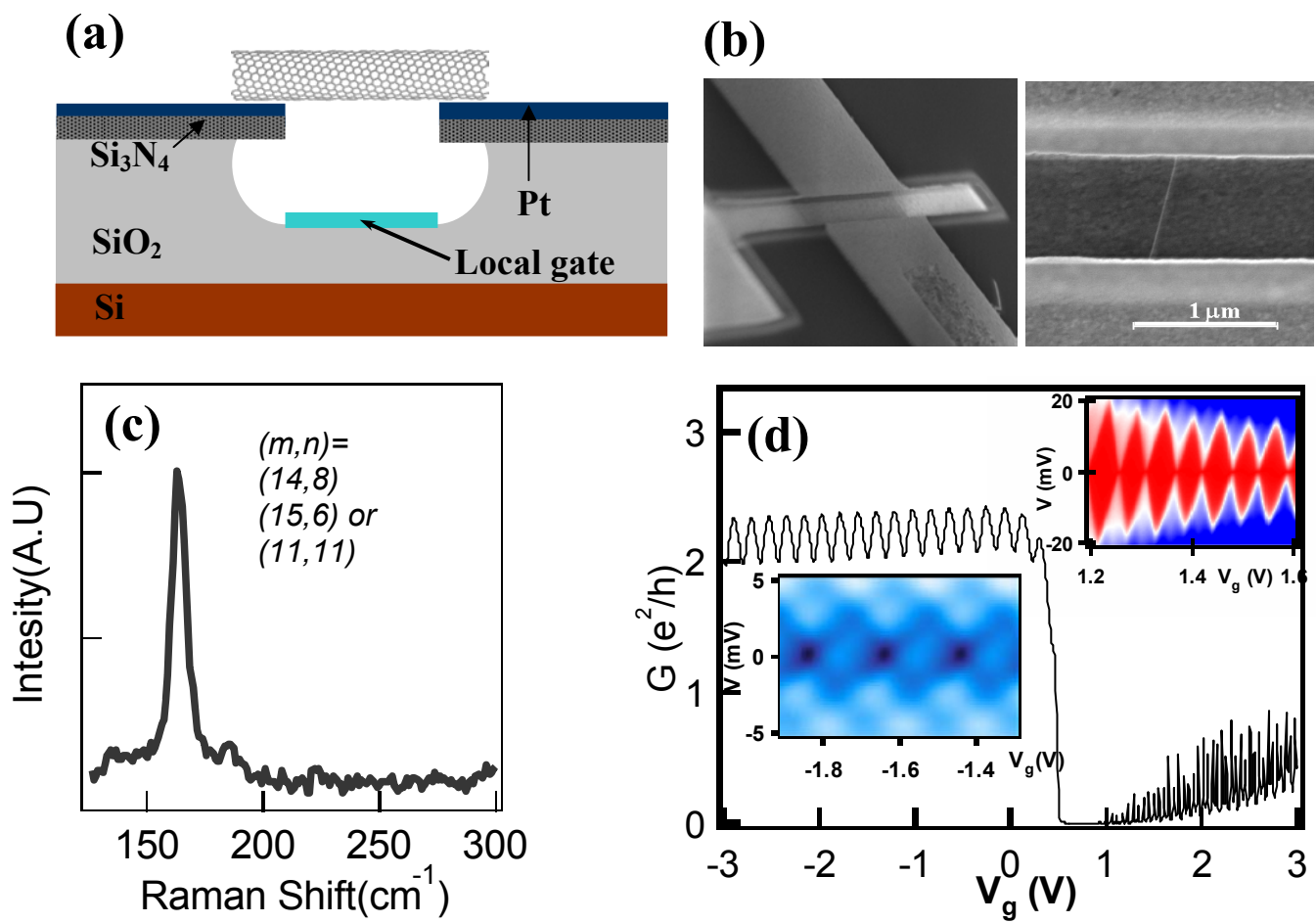

Fig. 1

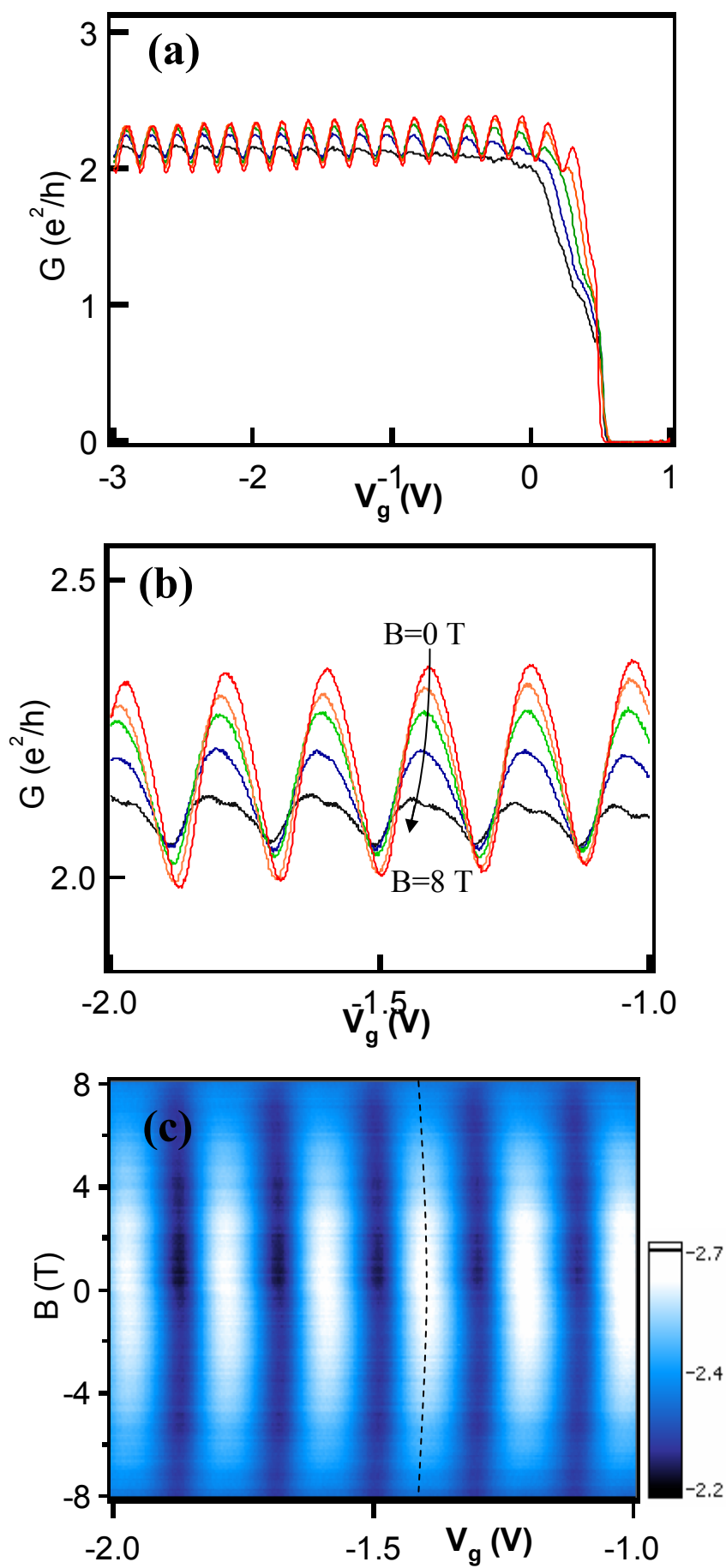

**(a)**

$G$ ($e^2/h$)

$V_g^{-1}$ (V)

**(b)**

$G$ ($e^2/h$)

B=0 T

B=8 T

$\bar{V}_g^{1.5}$ (V)

**(c)**

B (T)

$V_g$ (V)

Fig. 2

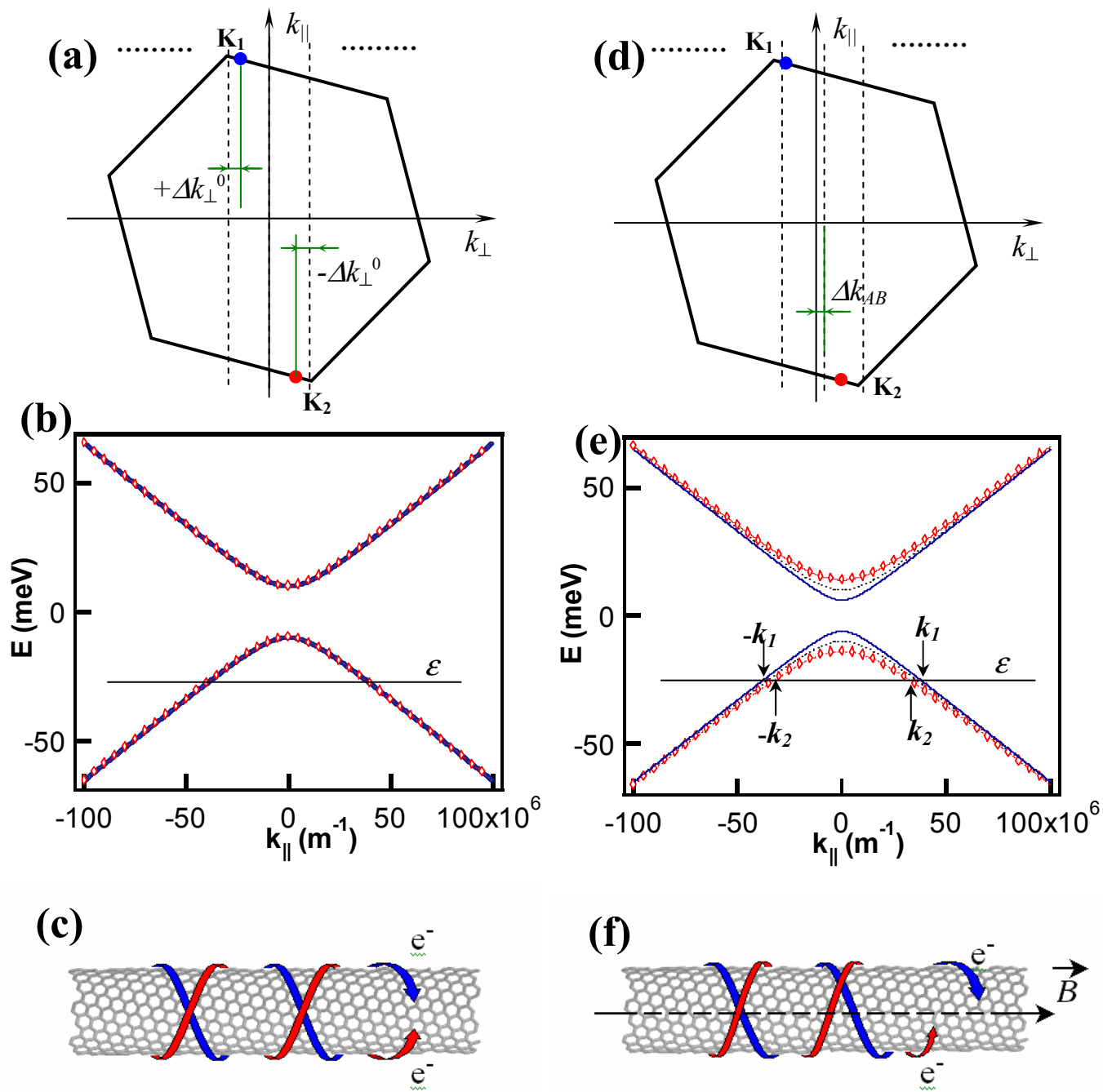

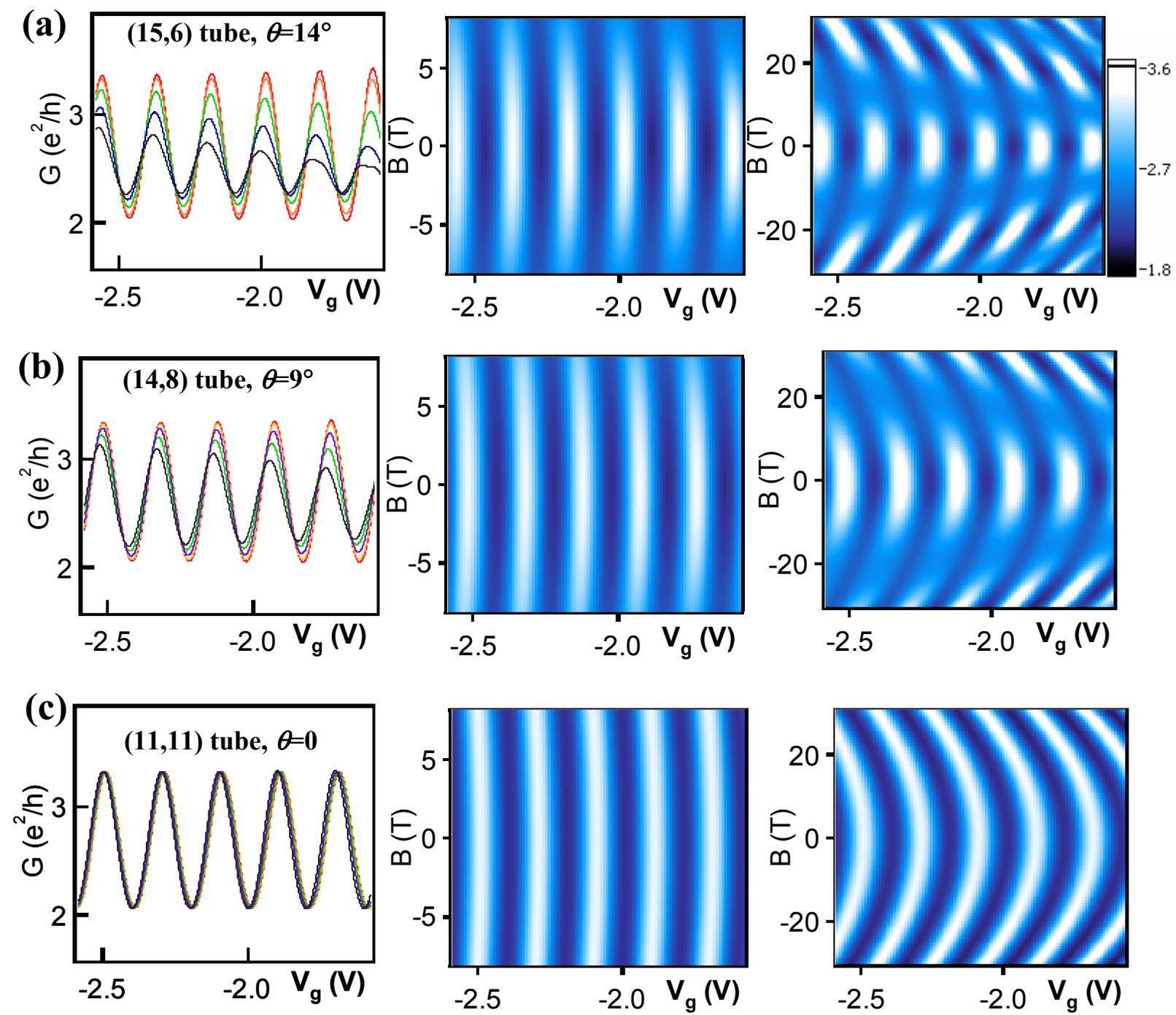

**(a)** (15,6) tube, θ=14°

**(b)** (14,8) tube, θ=9°

**(c)** (11,11) tube, θ=0

Fig. 4